\UseRawInputEncoding
\documentclass[pdflatex,sn-mathphys-num]{sn-jnl}% Math and Physical Sciences
\usepackage{chngcntr}
\counterwithout{table}{section} % If you want to keep table numbers independent of sections
\counterwithout{figure}{section} % Same for figures
\usepackage{graphicx}%
\usepackage[table]{xcolor}
\usepackage{amsmath,amssymb,amsfonts}%
\usepackage{amsthm}%
\usepackage{mathrsfs}%
\usepackage[title]{appendix}%
\usepackage{textcomp}%
\usepackage{manyfoot}%
\usepackage{booktabs}%
\usepackage{algorithm}%
\usepackage{algorithmicx}%
\usepackage{algpseudocode}%
\usepackage{listings}%
\usepackage{hyphenat}
\usepackage{float}
\usepackage{tabularx}
\usepackage{multirow}

\begin{document}
\title[Instruction-Based Fine-tuning of Open-Source LLMs for Predicting Customer Purchase Behaviors]{Instruction-Based Fine-tuning of Open-Source LLMs for Predicting Customer Purchase Behaviors}


\author[1]{\fnm{Halil Ibrahim} \sur{Ergul}}\email{halil.ergul@sabanciuniv.edu}
\equalcont{These authors contributed equally to this work.}

\author[1]{\fnm{Selim} \sur{Balcisoy}}\email{selim.balcisoy@sabanciuniv.edu}
\equalcont{These authors contributed equally to this work.}

\author[2]{\fnm{Burcin} \sur{Bozkaya}}\email{bbozkaya@sabanciuniv.edu}
\equalcont{These authors contributed equally to this work.}

\affil[1]{\orgdiv{Faculty of Engineering and Natural Sciences}, \orgname{Sabanci University}, \city{Istanbul}, \country{Turkiye}}

\affil[2]{\orgdiv{School of Management}, \orgname{Sabanci University}, \city{Istanbul}, \country{Turkiye}}


\abstract{In this study, the performance of various predictive models, including probabilistic baseline, CNN, LSTM, and fine-tuned LLMs, in forecasting merchant categories from financial transaction data have been evaluated. Utilizing datasets from Bank A for training and Bank B for testing, the superior predictive capabilities of the fine-tuned Mistral Instruct model, which was trained using customer data converted into natural language format have been demonstrated. The methodology of this study involves instruction fine-tuning Mistral via LoRA (Low-Rank Adaptation of Large Language Models) to adapt its vast pre-trained knowledge to the specific domain of financial transactions. The Mistral model significantly outperforms traditional sequential models, achieving higher F1 scores in the three key merchant categories of bank transaction data—grocery, clothing, and gas stations— that is crucial for targeted marketing campaigns. This performance is attributed to the model's enhanced semantic understanding and adaptability which enables it to better manage minority classes and predict transaction categories with greater accuracy. These findings highlight the potential of LLMs in predicting human behavior and revolutionizing financial decision-making processes}



\maketitle

\section{Introduction}
Large Language Models (LLMs) have demonstrated extraordinary proficiency in generating text that closely mimics human language, as well as excelling in a wide array of tasks, such as Natural Language Processing, Information Retrieval, and recommendation \cite{zhao2023survey} \cite{chowdhery2022palm} \cite{liang2023holistic} \cite{wei2022emergent}. More importantly, these large models have also proven to be successful in improving techniques commonly used to study and model human behavior \cite{Bail2024Can}. Extensive research has highlighted their ability to encode rich knowledge and exhibit compositional generalization, which allows them to apply learned concepts to novel scenarios effectively. When given appropriate instructions, these models can leverage their vast knowledge bases to solve previously unseen tasks and achieve remarkable levels of performance \cite{zhang2024instructiontuninglargelanguage} \cite{wang2024surveydataselectionllm}. The versatility and depth of understanding that LLMs offer make them particularly well-suited to addressing complex challenges that require not only strong generalization capabilities but also an in-depth grasp of contextual and semantic intricacies. These capabilities position LLMs as transformative tools with the potential to significantly advance various domains, especially for human behavior modeling problems that demand both extensive knowledge and adaptable reasoning.

One such domain that potentially stands to benefit greatly from the advanced capabilities of LLMs is financial prediction. In the context of financial transactions, predicting user preferences based on historical data is a complex sequential task. This complexity arises from the diverse and imbalanced nature of transaction categories as well as different characteristics of customers' demographic features, which necessitates models that can accurately capture and interpret subtle patterns in user behavior. Conventional neural network models that are used for this next category prediction task, such as Convolutional Neural Networks (CNNs) and Long Short-Term Memory (LSTM) networks, have been employed to deal with this problem with varying degrees of success \cite{Shikov2019-xu}. Despite significant advancements in the field, the domain of sequential prediction of human behavior, particularly regarding next merchant category prediction using transactional customer data, remains relatively unexplored. These sequential models, while effective in certain contexts, often struggle with capturing the patterns and semantic nuances inherent in transaction data, particularly when dealing with real-world imbalanced datasets where minority classes are underrepresented.

\begin{figure*}%[tbhp]
\centering
\includegraphics[width=\linewidth]{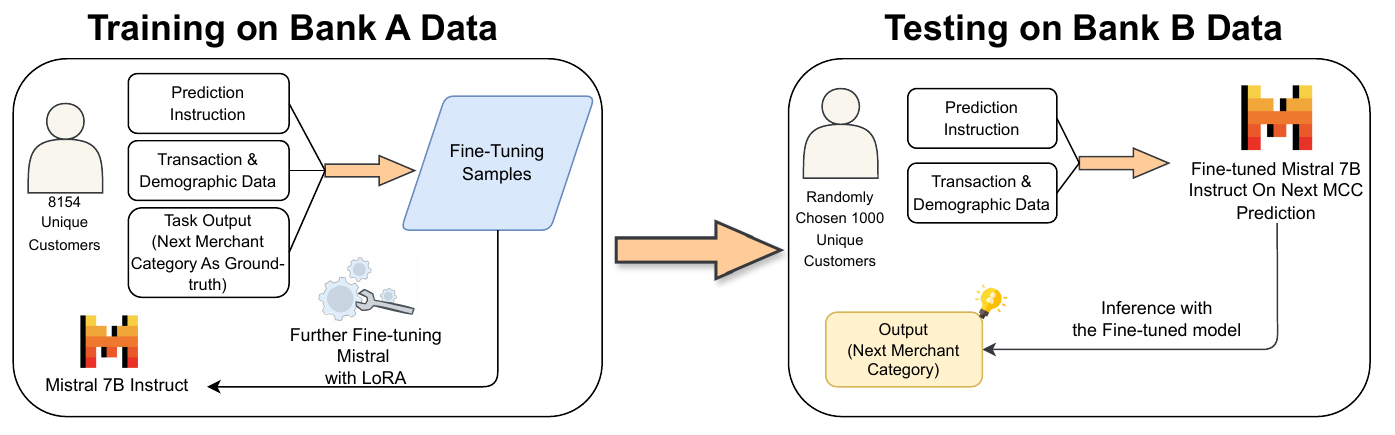}
\caption{An overview of the complete process for fine-tuning and evaluating the Mistral 7B Instruct model on predicting next merchant categories, utilizing distinct datasets from Bank A for training and Bank B for testing to evaluate generalizability}
\label{fig:model-illustration}
\end{figure*}

This study leverages the broader capability of LLMs to predict human behavior by applying this to the specific context of financial transactions, where accurate predictions are particularly valuable. By addressing the complex problem of predicting customer purchase behaviors, it has been demonstrated that LLMs solve this issue effectively and with greater efficiency and less modeling effort compared to deep learning models. This approach not only highlights the efficacy of LLMs in understanding and predicting human behavior but also emphasizes their applicability in critical financial contexts. Specifically, the aim of this study is to come up with a novel methodology for predicting the next purchase merchant categories of customers by fine-tuning an open-source LLM and comparing its performance with deep-learning based sequential models. To the best of our knowledge, this is the first study to employ a similar methodology for this specific prediction task with the purpose of human behavior modeling, which distinguishes this work from existing literature. By leveraging datasets from two different banks (Bank A and Bank B), models have been trained on data from Bank A and test their generalization capabilities on previously unseen data from Bank B. The experimental evaluation is designed to illustrate the performance of these models across different sequence lengths of user preferences within a multi-class classification framework.

Figure \ref{fig:model-illustration} illustrates the main case study of this paper that contains the model training and testing pipeline designed for predicting the next merchant category based on historical customer transactions and demographic data. The training phase utilizes data from 8,154 unique customers of Bank A, comprising prediction instructions and task outputs with the next category as the ground truth. Fine-tuning samples are derived from this dataset to further fine-tune the Mistral 7B Instruct model. In the testing phase, the model is validated using data from 1,000 randomly chosen unique customers from Bank B. This data, paired with prediction instructions, is used to test the inference capabilities of the fine-tuned model, which then outputs predictions for the next merchant category. The pipeline underscores the importance of fine-tuning and cross-validation across different datasets to ensure the model's robustness and predictive accuracy in varied contexts. Notably, the dataset from Bank A spans a distinct time period compared to the dataset from Bank B, which ensures temporal independence between the training and testing phases.

This study addresses several critical research questions that have yet to be thoroughly explored. Firstly, it has been investigated whether a pre-trained LLM can be effectively fine-tuned on transactional tabular data reformatted into personalized instructions for merchant category prediction tasks. Secondly, it has been evaluated the performance of the fine-tuned model against traditional deep-learning-based sequential prediction models. Lastly, the model's performance across various sequential lengths that were not originally part of the training data has been assessed. By answering these questions, this research aims to provide comprehensive insights into the capabilities and limitations of fine-tuned LLMs in the context of financial transaction prediction.

\section{Results}
\begin{table*}[]
\caption{Overall Evaluation Scores Across All Merchant Categories}
\label{tab:my-table}
\resizebox{\columnwidth}{!}{%
\begin{tabular}{@{}|c|c|c|c|c|c|c|@{}}
\toprule
\textbf{Model} &
  \textbf{Dataset} &
  \textbf{Sequence Length} &
  \textbf{Accuracy} &
  \textbf{Precision} &
  \textbf{Recall} &
  \textbf{F1 (weighted)} \\ \midrule
\begin{tabular}[c]{@{}c@{}}Mistral Instruct 7b v.2 \\ (Raw Model)\end{tabular} &
  \cellcolor[HTML]{FFFFFF}Bank B &
  last-9 &
  0.51 &
  0.50 &
  0.50 &
  0.49 \\ \midrule
 &
  \cellcolor[HTML]{FFFFFF} &
  last-9 &
  0.42 &
  0.50 &
  0.42 &
  0.43 \\ \cmidrule(l){3-7} 
 &
  \cellcolor[HTML]{FFFFFF} &
  last-4 &
  0.43 &
  0.52 &
  0.43 &
  0.43 \\ \cmidrule(l){3-7} 
 &
  \cellcolor[HTML]{FFFFFF} &
  last-7 &
  0.42 &
  0.50 &
  0.42 &
  0.42 \\ \cmidrule(l){3-7} 
\multirow{-4}{*}{\begin{tabular}[c]{@{}c@{}}Baseline \\ (Averaging)\end{tabular}} &
  \multirow{-4}{*}{\cellcolor[HTML]{FFFFFF}Bank B} &
  last-14 &
  0.43 &
  0.49 &
  0.43 &
  0.44 \\ \midrule
 &
   &
  last-9 &
  0.65 &
  0.61 &
  0.65 &
  0.62 \\ \cmidrule(l){3-7} 
 &
   &
  last-4 &
  0.63 &
  0.60 &
  0.63 &
  0.60 \\ \cmidrule(l){3-7} 
 &
   &
  last-7 &
  0.64 &
  0.61 &
  0.64 &
  0.61 \\ \cmidrule(l){3-7} 
\multirow{-4}{*}{\begin{tabular}[c]{@{}c@{}}CNN \\ (trained on Bank A)\end{tabular}} &
  \multirow{-4}{*}{Bank B} &
  last-14 &
  0.66 &
  0.62 &
  0.66 &
  0.62 \\ \midrule
 &
   &
  last-9 &
  0.62 &
  0.60 &
  0.62 &
  0.60 \\ \cmidrule(l){3-7} 
 &
   &
  last-4 &
  \textbf{0.63} &
  0.59 &
  \textbf{0.62} &
  0.58 \\ \cmidrule(l){3-7} 
 &
   &
  last-7 &
  \textbf{0.64} &
  0.61 &
  \textbf{0.64} &
  0.60 \\ \cmidrule(l){3-7} 
\multirow{-4}{*}{\begin{tabular}[c]{@{}c@{}}LSTM \\ (trained on Bank A)\end{tabular}} &
  \multirow{-4}{*}{Bank B} &
  last-14 &
  \textbf{0.63} &
  0.62 &
  \textbf{0.63} &
  0.61 \\ \midrule
 &
   &
  last-9 &
  \textbf{0.66} &
  \textbf{0.67} &
  \textbf{0.66} &
  \textbf{0.66} \\ \cmidrule(l){3-7} 
 &
   &
  last-4 &
  0.61 &
  \textbf{0.69} &
  0.60 &
  \textbf{0.62} \\ \cmidrule(l){3-7} 
 &
   &
  last-7 &
  0.60 &
  \textbf{0.64} &
  0.62 &
  \textbf{0.63} \\ \cmidrule(l){3-7} 
\multirow{-4}{*}{\begin{tabular}[c]{@{}c@{}}Mistral Instruct 7b v.2 \\ (trained on Bank A)\end{tabular}} &
  \multirow{-4}{*}{Bank B} &
  last-14 &
  0.62 &
  \textbf{0.66} &
  0.62 &
  \textbf{0.63} \\ \bottomrule
\end{tabular}%
}
\end{table*}

The results of the experimental evaluation are organized to illustrate the performance of various models across a series of sequence lengths and to show the ability of each model to predict user preferences for merchant categories in a multiclass classification framework. The dataset utilized for training and validation purposes, Bank A, served as the basis for training several models. Bank B, an entirely separate dataset not used for training, provided an additional layer of testing to assess the models' generalization capabilities. 

Table \ref{tab:my-table} shows the results of baseline, neural networks (CNN and LSTM), and the fine-tuned LLM model (for overall results of all fine-tuned LLM models see Supplementary \textbf{Table 1}). The bold scores were depicted by comparing LSTM and CNN with Mistral model. The “Dataset” column indicates the specific dataset used for making predictions or inferences with each model. Sequence length values demonstrate how many transactions of customers were used as input for the models to get a transaction category prediction. For instance, “last-7” means that the last seven transactions of customers were given as input for a model to get an inference regarding their 8th transaction category. It's important to note that all models, with the exception of the baseline averaging model which does not require any training, were trained using the last nine transactions (“last-9”) from customers in Bank A dataset and the 10th transaction category used as a ground truth or label.

\subsection{Overall Results in Generalization to Bank B Data}

The Baseline model shows consistently lower performance metrics across accuracy, precision, recall, and F1 scores when compared to neural network approaches and the fine-tuned LLM. Specifically, the model's accuracy hovers around 0.42 to 0.43 across different sequence lengths, with precision and recall metrics similarly ranging between 0.49 to 0.52 and 0.42 to 0.43, respectively. The F1 score, which considers both precision and recall, remains relatively low, ranging from 0.42 to 0.44. The Baseline model's performance, while providing a necessary benchmark for comparison, highlights its inadequacy for complex predictive tasks such as merchant category forecasting. Its lower performance metrics relative to other evaluated models reinforce the importance of adopting more sophisticated models that can better capture and analyze the intricacies of financial transaction data.

The analysis of the Convolutional Neural Network (CNN) and Long Short-Term Memory (LSTM) models, both trained on Bank A's data and tested on Bank B, demonstrates their relative performance metrics in predicting the next merchant categories. The CNN model exhibits moderate effectiveness across various sequence lengths, with the best performance observed at the last-14 transactions, achieving an accuracy of 0.66, a precision of 0.62, and a recall of 0.66, resulting in an F1 score of 0.62. When considering shorter sequence lengths, the performance slightly decreases, with the model demonstrating accuracy and F1 score around 0.63-0.65 and 0.60-0.62 respectively. These metrics suggest that while the CNN can handle different contexts, its predictive capabilities are not strongly enhanced by extending the sequence length.

The LSTM model, known for its ability to capture temporal dependencies, shows a consistent pattern across different transaction sequence lengths. The performance peaks with a sequence length of the last-14 transactions, where it achieves an accuracy of 0.63, a precision of 0.62, and a recall of 0.63, yielding an F1 score of 0.61. Similar to the CNN, the LSTM model does not demonstrate substantial improvement as the sequence length increases, with metrics revolving around 0.62 for accuracy and 0.60 for the F1 score across different contexts. 

The fine-tuned Mistral Instruct model achieves a weighted F1 score of 0.66 when evaluating the last-9 transactions, which is notably higher than the scores obtained by both the CNN and LSTM models under the same conditions (0.62 and 0.60 respectively). These superior F1 scores across all sequence lengths indicate that the fine-tuned Mistral Instruct model not only predicts more accurately but also maintains a balanced sensitivity and precision, which is crucial in handling class-imbalanced datasets effectively. When compared to the other baseline models, the fine-tuned Mistral Instruct stands out, particularly in terms of the F1 score, which is crucial for evaluating performance in an imbalanced dataset context. The model not only outperforms the CNN and LSTM in this metric but also exhibits a higher consistency across different sequence lengths. This highlights its adaptability and overall superior predictive power in this specific task.

% \usepackage{booktabs}
% \usepackage{multirow}
% \usepackage{graphicx}
\begin{table*}[]
\caption{Class-wise F1 Scores for LSTM, CNN, and Mistral LLM Models}
\label{tab:my-table2}
\resizebox{\textwidth}{!}{%
\begin{tabular}{@{}|c|c|c|c|c|c|@{}}
\toprule
\textbf{Model} &
  \textbf{Dataset} &
  \textbf{Sequence Length} &
  \textbf{Clothing} &
  \textbf{Gas Stations} &
  \textbf{Grocery} \\ \midrule
\multirow{4}{*}{CNN}  & \multirow{4}{*}{Bank B} & last-9  & 0.004          & 0.222          & 0.419          \\ \cmidrule(l){3-6} 
                      &                         & last-4  & 0.107          & 0.185          & 0.417          \\ \cmidrule(l){3-6} 
                      &                         & last-7  & 0.017          & 0.209          & 0.422          \\ \cmidrule(l){3-6} 
                      &                         & last-14 & 0.000          & 0.222          & 0.410          \\ \midrule
\multirow{4}{*}{LSTM} & \multirow{4}{*}{Bank B} & last-9  & 0.001          & 0.230          & 0.450          \\ \cmidrule(l){3-6} 
                      &                         & last-4  & 0.006          & 0.200          & 0.346          \\ \cmidrule(l){3-6} 
                      &                         & last-7  & 0.003          & 0.223          & 0.448          \\ \cmidrule(l){3-6} 
                      &                         & last-14 & 0.000          & 0.244          & 0.474          \\ \midrule
\multirow{4}{*}{\begin{tabular}[c]{@{}c@{}}Mistral Instruct \\ 7b v.2\end{tabular}} &
  \multirow{4}{*}{Bank B} &
  last-9 &
  \textbf{0.620} &
  \textbf{0.400} &
  \textbf{0.590} \\ \cmidrule(l){3-6} 
                      &                         & last-4  & \textbf{0.480} & \textbf{0.470} & \textbf{0.560} \\ \cmidrule(l){3-6} 
                      &                         & last-7  & \textbf{0.120} & \textbf{0.500} & \textbf{0.520} \\ \cmidrule(l){3-6} 
                      &                         & last-14 & \textbf{0.220} & \textbf{0.480} & \textbf{0.550} \\ \bottomrule
\end{tabular}%
}
\end{table*}

\subsection{Class-Specific Performance}
In Table \ref{tab:my-table2}, a detailed analysis of the classification performance across individual merchant categories for three models that are all trained on Bank A dataset is presented (for class-wise scores across all categories including “Other” of all fine-tuned models see Supplementary \textbf{Table 2}). The distribution of classes in the training set from Bank A (also similar to Bank B data) was notably skewed, with the grocery category at 31.3\%, 11.9\% to gas stations, and 11.2\% to clothing. The remaining 45.5\% of transactions pertain to categories different than grocery, clothing, and gas stations. This class imbalance presents a challenge for predictive models, particularly for minority classes that are underrepresented in the data.

The evaluation of class-specific F1 scores offers a comprehensive view of the performance disparities between the fine-tuned Mistral Instruct 7b v.2 model and traditional sequential models such as CNNs and LSTMs. This analysis is critical as it illustrates the refined capability of the Mistral model to handle minority classes with a higher level of accuracy and efficiency, compared to the more generic treatment of classes by the other models.

In the 'Clothing' category, the Mistral Instruct model demonstrates a pronounced superiority with an F1 score of 0.620 when analyzing the last-9 transaction sequences. This score is significantly higher than those achieved by the CNN and LSTM models, which reach only 0.107 and 0.006, respectively. Such a stark difference underscores the fine-tuned model's approach to semantic learning and prediction in categories that represent a smaller portion of the dataset, specifically the 11.2\% comprising clothing transactions in Bank A's data.

Similarly, for the “Gas Stations” category, the Mistral model achieves its peak performance with an F1 score of 0.500 using the last-7 transactions. This outstrips the CNN and LSTM models, which max out at F1 scores of 0.222 and 0.244. The improved performance in this category, which constitutes 11.9\% of the transactions in the training dataset, highlights the model's effective adaptation to categories characterized by lower frequency yet distinct spending patterns.

The “Grocery” category reveals the model's exceptional ability to manage relatively more frequent transaction classes but still not the majority, with an impressive F1 score of 0.59 from the last-9 transaction sequences. This compares favorably with the best scores from CNN and LSTM, which are 0.422 and 0.474, respectively. Given that grocery transactions make up 31.3\% of Bank A's data, this result emphasizes the Mistral model's robust predictive capabilities to handle dominant classes effectively as well.

These findings not only reinforce the effectiveness of the Mistral model in dealing with class imbalances but also highlight its suitability for complex predictive tasks in real-world financial environments. The model's ability to excel in predicting minority class categories, where traditional models struggle, sets a new standard for the usage of LLMs in the domain of financial transactions. The detailed class-specific performance thus underlines the transformative impact of fine-tuning large language models on domain-specific tasks.

The fine-tuned Mistral Instruct 7b v.2 model exhibits remarkable consistency across all categories. This demonstrates its stability and reliability in performance when compared to traditional sequential models like CNNs and LSTMs. This consistency is not only evident in the superior scores it achieves in the minority categories but also in its robust performance across varying transaction sequences and class distributions. Unlike the CNN and LSTM models, which show considerable fluctuations in their class-specific F1 scores, the Mistral model maintains a more uniform performance spectrum. For instance, while the F1 scores for CNN and LSTM vary widely from nearly zero in some categories to higher values in others, the Mistral model's scores remain notably higher and more stable. This shows more stability in reducing the performance variance between the best and worst-performing categories. This consistent performance is crucial for applications that require dependable and predictable model behavior across diverse and potentially imbalanced datasets.

\section{Discussion}
One of the most notable outcomes of this study is the demonstrated superiority of the fine-tuned Mistral Instruct model over traditional sequential models especially in class-wise scores. This superiority is evident across several metrics and transaction categories. The fine-tuning process enables the model to adapt its vast pre-trained knowledge to the specific domain of financial transactions. This adaptation involves aligning the model’s parameters with the unique characteristics of the transaction data, including understanding the semantics within the merchant categories and the contextual information surrounding each transaction.

The semantic understanding capability of LLMs plays a crucial role in their performance in modeling human behavior. When dealing with merchant categories like 'Clothing' or 'Gas Stations', the LLM can  “understand” the meaning and context associated with these categories and can potentially relate these to the demographic data of customers. This deeper semantic understanding allows the LLM to think beyond mere sequential logic and enables it to consider the broader implications and associations of each category. In contrast, traditional models like CNNs and LSTMs treat these categories as sequential numbers without inherent meaning, which limits their ability to capture complex patterns and contextual nuances.

Another critical aspect of this study is the handling of class imbalances. Transaction data, by nature, often exhibit significant imbalances across different categories. Traditional models like CNNs and LSTMs typically struggle with such imbalances and often overfitting to the majority class and underperforming in minority classes. The Mistral model, however, showed a remarkable ability to handle these imbalances. It maintained high performance across both majority and minority classes. This balanced performance is crucial for practical applications where accurate predictions across all categories are necessary. The model’s ability to handle minority classes with higher precision, as seen in the “Clothing” and “Gas Stations” categories, underscores its potential in real-world financial environments where class imbalances are a common challenge. It is also important to note that to preserve the original behavioral transaction sequences of customers, categories that are not among the three “Grocery”, “Clothing”, and “Gas Stations” were labeled as “Other” during both training and testing data.

Lastly, the results suggest that in three out of the four major categories—'Grocery', 'Clothing', and 'Gas Stations'—the LLM's performance significantly outpaced that of traditional neural net models. These findings are particularly relevant for banks and financial institutions, which often focus their credit card reward programs and marketing efforts on popular spending categories such as groceries, clothing, and gas stations. By leveraging the predictive power of fine-tuned LLMs, banks can develop more effective and personalized marketing strategies. This can translate into better customer experiences and increased usage of bank services, which highlights the practical benefits of integrating advanced machine learning models into financial decision-making processes.

\section{Methods}
\subsection{Raw data}
A major financial institution in an OECD country donated two deidentified samplings of their data (called as \textbf{Bank A} throughout this study) that were collected over the period between July 2014 and July 2015 \cite{Kaya2018Behavioral}.  The dataset presented comprises a range of attributes that offer a multifaceted view of customer transactions. Each entry includes a masked customer identifier to maintain privacy while allowing for the tracking of individual customer activities. 

While Bank A data was used both in the training and testing phases, another dataset from a different bank (called as \textbf{Bank B} throughout this study) was only used for testing purposes to see how well the trained models on Bank A will be performing on Bank B. Bank B dataset includes tens of thousands of individual accounts, which represent about 10\% of the total customer accounts from a data warehouse of a major financial institution in an OECD country \cite{Singh2015Money}. Each customer account contains data on all credit card transactions made for purchases over a three-month period in 2013.

The transaction date and time records when the purchase occurred, and the transaction amount captures the value of the purchase. Customer demographics are detailed through columns indicating gender and marital status, while socioeconomic variables are represented by fields detailing education level, job description, and income. The most important attribute and the label (ground truth) attribute is the merchant category of the transaction, which informs the type of goods or services purchased. Additionally, one generated category out of raw income information is income group to categorize raw income into segments of low, middle, and high-income groups. This structured data serves as the foundation for preparing the fine-tuning data, which will predict future customer purchase categories based on transactional history.

\subsection{Raw Data Preprocessing}
In preparing the transactional data for the fine-tuning of a language model, a series of preprocessing decisions were implemented to ensure the integrity and quality of the dataset. The preprocessing steps included filtering only merchant categories to the three most frequently purchased categories. This decision was driven by the need to mitigate class imbalance, particularly as certain categories like insurance were significantly less represented. The three categories selected were Grocery Stores, Clothing Stores, and Gas Stations, representing the top tiers of transaction activity within the dataset. For the purpose of preserving the real-life transaction sequence and behaviors of customers, the merchant categories outside of these top three categories were labeled as “Other” and used in training as well as testing and treated as another category.

Users lacking complete demographic information were excluded to ensure a comprehensive dataset vital for training. The data cleansing process involved removing users with fewer than ten transactions across at least two distinct categories to eliminate insufficient histories that could skew predictions and to reduce noise from users predominantly purchasing from a single category like truck drivers who constantly make purchases from gas stations.

Through these preprocessing steps, the dataset was refined from an initial 10,000 users with 298 categories and over a million transactions to a more focused cohort of 8,154 users with transactions exclusively within the four targeted categories (Grocery, Clothing, Gas Stations, and Other), amounting to 1,123,445 transactions. These measures were essential in ensuring that the fine-tuning process of the language model would be conducted on high-quality and relevant data.

\subsection{Finetune Dataset Preparation and Formatting}
In transforming the customer data from its original tabular representation into a format suitable for instruction tuning of a language model, a systematic approach was adopted to articulate each customer's data in natural language. The methodology is based on a four\hyp{}step instruction tuning process to prepare the data for the model \cite{ouyang2022training}\cite{Bao_2023}. The objective is to enable the model to predict the 10th merchant category purchase for a customer, as demonstrated in an illustrative example referred to as Table \ref{tab:my-table3}.

\begin{table}[ht!]
\centering
\caption{A sample format created from the raw transaction dataset for finetuning.}
\label{tab:my-table3}
\begin{tabularx}{\columnwidth}{X}
\toprule
\textbf{Instruction Input} \\
\midrule
\textbf{Task Instruction:} Based on my demographic details and historical transaction data provided below, predict my next purchase category. \\
\textbf{Task Input:} I am 1695432. I am 48 years old, married male, secondary school graduate, and I am working as a private employee. In terms of my income state, I belong to the high-income group. Recently, I made 9 transactions. In these transactions, I have spent a total of \$560.61 dollars. I bought items from the following categories, chronologically: Grocery, Grocery, Grocery, Other, Clothing, Other, Clothing, Clothing, Clothing. I bought from these categories on the following dates, chronologically: 2015\hyp{}03\hyp{}28, 2015\hyp{}04\hyp{}01, 2015\hyp{}04\hyp{}15, 2015\hyp{}05\hyp{}01, 2015\hyp{}05\hyp{}27, 2015\hyp{}06\hyp{}04, 2015\hyp{}06\hyp{}04, 2015\hyp{}06\hyp{}04, 2015\hyp{}06\hyp{}08. I spent the following money for these items, chronologically: \$39.82, \$47.25, \$27.81, \$124.97, \$105.97, \$24.95, \$49.99, \$99.95, \$39.90.
 \\
\textbf{Instruction Output} \\
\midrule
\textbf{Task Output:} Gas stations. \\
\bottomrule
\end{tabularx}
\end{table}

The first step involves defining the task through a “Task Instruction”, which is presented in natural language and provides a concise yet comprehensive description of the objective, including the specific nature of the prediction to be made. The second step is the formulation of the “Task Input” and “Task Output” in natural language, where the “Task Input” encapsulates the customer’s demographic information and historical transaction data, while the “Task Output” signifies the anticipated future transaction category in this study.

In the third step, the “Task Instruction” and “Task Input” are synthesized to create the “Instruction Input”, which is a narrative embodying both the task at hand and the contextual data pertinent to the customer. The “Task Output” is then established as the “Instruction Output”, representing the expected model prediction. This “Instruction Input” and “Instruction Output” pair forms the foundation for each tuning sample.

The final step involves the actual instruction tuning on large language models (LLMs), employing the crafted pairs of “Instruction Input” and “Instruction Output” to train the model. Through this training, the model learns to anticipate the subsequent transaction category—a prediction of the 10th purchase based on the customer's demographic and transactional background.

An example of this methodology applied to a single customer is provided in the aforementioned Table \ref{tab:my-table3}. This example demonstrates the narrative structure adopted to present the customer’s profile and transaction history, followed by the model's anticipated classification of the next merchant category. This approach ensures that each customer’s data is transformed into a detailed narrative format to prepare the dataset for fine\hyp{}tuning the language model to predict future customer transactions.

\subsection{Finetuning Framework}
In this study, the finetuning framework deployed leveraged the LoRA (Low-Rank Adaptation of Large Language Models) technique accessed through the Parameter\hyp{}Efficient FineTuning (PEFT) library on the Hugging Face platform, together with the SFFtrainer library designed for Supervised FineTuning (SFT) \cite{peft} \cite{von_Werra_TRL_Transformer_Reinforcement}. The dataset structured for this purpose contains natural language representations of customer profiles and their transaction histories, with the 10th merchant category purchase serving as the label for supervised learning.

LoRA’s methodology is centered on improving the fine\hyp{}tuning efficiency of language models \cite{hu2021lora}. Rather than adjusting the entire set of weights in the pre\hyp{}trained model, LoRA introduces a pair of smaller matrices. These matrices, known as update matrices, undergo training to capture the new information presented in the fine\hyp{}tuning dataset. This is achieved through a process called low\hyp{}rank decomposition. The core idea is to maintain the pre\hyp{}trained weights unaltered—effectively freezing them—thus preserving the knowledge previously acquired by the model while still allowing it to adapt to new tasks. This tuning approach has the following objective function:

\begin{equation}
\max_{\Theta} \sum_{(x,y) \in Z} \sum_{t=1}^{|y|} \log \left( P_{\phi+\Theta} (y_t \mid x, y_{<t}) \right).
\end{equation}

The learning objective function is characterized by the maximization of the log probability, which is the sum of the log probabilities of each token \( y_t \) given the input sequence \( x \) and the preceding tokens \( y_{<t} \), across all input-output pairs in the training set \( Z \). This probability is a function of both the original frozen parameters of the model \( \phi \) and the adjustments \( \Theta \) made by the low-rank parameter updates. In this context, \( x \) represents the input sequence, \( y \) is the corresponding output sequence, \( y_t \) is the \( t \)-th token in the output sequence, \( y_{<t} \) denotes all tokens preceding \( y_t \) in the output sequence, \( Z \) is the training set consisting of input-output pairs \( (x, y) \), \( \phi \) are the original (frozen) parameters of the model, and \( \Theta \) are the low-rank parameters introduced to adjust the model. The objective function seeks to optimize the parameter set \( \Theta \) such that the log probability of the output sequence \( y \) given the input sequence \( x \) and the previously generated tokens \( y_{<t} \) is maximized. This method leverages both the inherent knowledge in the frozen parameters \( \phi \) and the fine-tuning capacity introduced through the low-rank parameters \( \Theta \).

By keeping the pre\hyp{}trained weights frozen and introducing only a small number of new trainable parameters, the intrinsic information within the language model is leveraged efficiently. Thus, the fine\hyp{}tuning process incorporates additional information without overwriting the valuable insights already encapsulated within the pre\hyp{}existing model structure.

\subsection{LLM Selection And Experiments}
The chosen LLMs are distinguished by their pre-existing instruction tuning on task-oriented activities, their accessibility as open-source weights, their documentation quality, their commendable performance in benchmarks pertinent to natural language processing tasks, and their recognition within the open-source LLM community. Each model was fine-tuned using a dataset comprising 8154 unique customers, representing 90\% of the total dataset. For this training, a single Nvidia RTX 3090 with 24 GB RAM was used.

The first two models subjected to fine-tuning via the Alpaca method were the LLaMA-2-Chat 7B and LLaMA-3-8B instruct, auto-regressive language models utilizing an optimized transformer architecture \cite{touvron2023llama}\cite{alpaca}\cite{wang2023selfinstruct}. This model has been further refined by Meta through both supervised fine-tuning (SFT) and reinforcement learning with human feedback (RLHF), with a focus on alignment with human preferences for helpfulness and safety. The architecture enhancements enable the model to engage in dialogue with improved responsiveness and ethical awareness.

The third and the most efficacious model in the fine-tuning experiments was the Mistral-7B-Instruct-v0.2 created by Mistral. This model is a fine-tuned iteration of the Mistral-7B-v0.2 and has been optimized using instruction tuning on publicly available datasets hosted on the Hugging Face platform \cite{jiang2023mistral}. Noteworthy is its superior performance over the LLaMA 2 and LLaMA 3 models as demonstrated in the Supplementary \textbf{Table 1} and \textbf{Table 2}. The Mistral-7B-Instruct-v0.2 employs Grouped-query attention (GQA) for accelerated inference and Sliding Window Attention (SWA) for efficiently managing longer sequences. Moreover, it utilizes a Byte-fallback BPE tokenizer, enhancing its capability to process and understand a diverse range of linguistic inputs.

Each model's architecture and training history contribute to its unique strengths in processing and predicting natural language patterns. The fine-tuning exercises conducted as part of this research were aimed at leveraging these strengths to predict the transactional behavior of customers, a task that necessitates a deep understanding of both language and human purchase behavior. The results from these experiments were expected to provide insights into the models' adaptability and performance in the specific context of merchant category prediction.

\subsection{Benchmark Models And Evaluation Strategy}
The task at hand is to predict user preference among four categories (in which one of them was labeled as 'Other'), which constitutes a multiclass classification problem. To this end, a suite of standard evaluation metrics is utilized, comprising accuracy, precision, recall, and the weighted F1 score. These metrics provide a holistic view of the model's performance by not only considering the proportion of correct predictions (accuracy) but also the models' ability to correctly identify positive cases (precision and recall) and a weighted measure of precision and recall (weighted F1 score).

For the testing phase as shown in Figure \ref{fig:model-illustration}, the fine-tuned (Mistral) and trained models (CNN and LSTM) were applied to data from Bank B, which is entirely separate and distinct from the training data (Bank A), including 1,000 randomly selected unique customers. This dataset is used to evaluate the model's performance and generalizability. The transaction history and demographic data for these customers are processed similarly to the training data. The fine-tuned Mistral 7B Instruct model predicts the next merchant category based on the provided inputs, and its output is compared against the actual 10th (or 8th, 15th, 5th depending on the sequence length to be tested) transaction's merchant category to assess models' accuracy on this unseen dataset. This testing phase is crucial as it demonstrates the models' ability to generalize and perform well on completely new data.

In the testing phase, various input sequence lengths were employed to rigorously assess the models' generalizability and to provide a comprehensive range of performance scores. Despite being trained on the last nine transactions, the models were tested with sequences of four, seven, and fourteen transactions. The first reason for this approach is to evaluate the models' adaptability to different data patterns that reflect real-world scenarios where transaction history may vary. The second reason is to generate multiple performance scores, which can show the models' strengths and limitations across various conditions. This provides a fuller picture of each model's predictive capabilities to ensure a more reliable and nuanced understanding of their performance.

To establish a baseline for comparison, the following models are employed with sequential modeling because the approach of this paper utilizes historical interactions to predict the subsequent interaction, similar to the sequential recommendation, following sequential models were considered:

Baseline Method (Averaging): The simplest model used for benchmarking is based on the historical average frequencies of transaction categories for each client. The prediction of a purchase in a specific category by a client is calculated using the formula:

\begin{align}
p_{ij} = \frac{1}{K_i} \sum_{k=1}^{K_i} I(n_{ijk} > 0)
\label{eq:probability_formula}
\end{align}

Here, $p_{ij}$ represents the probability of the $i$\hyp{}th client making a purchase in the $j$\hyp{}th category, $K_i$ denotes the total number of time periods considered, $n_{ijk}$ is the number of transactions for the $i$\hyp{}th client in the $j$\hyp{}th category at the $k$\hyp{}th time period, and $I$ is an indicator function that evaluates to 1 if the client has made one or more transactions in that category during the period.

The second model that has been trained for performance comparison was a simple LSTM architecture, with vectors representing encoded transaction categories serving as input. This LSTM is configured with a hidden state dimensionality of 128 and is trained to capture temporal transaction patterns using Backpropagation Through Time (BPTT) and a cross\hyp{}entropy loss function suited for classification tasks.

Lastly, a Convolutional Neural Network (CNN) model was trained. The network's input layer accepts concatenated vectors of encoded transaction categories, feeding into a straightforward CNN model with two 2D convolutional layers followed by a pooling layer. The network's architecture is completed with a final output layer that shares the same loss function as the LSTM, enabling direct performance comparisons between the recurrent and convolutional network strategies.

These benchmark models create a broad spectrum for comparison, ranging from the non\hyp{}ML averaging baseline to more complex neural architectures. The inclusion of these varied benchmarks allows for a well\hyp{}rounded evaluation of the fine\hyp{}tuned model's accuracy and reliability in predicting user transaction behaviors within the predefined merchant categories.

\section{Conclusion}
The primary motivation of this study was to explore the potential of LLMs in the domain of financial transaction prediction, particularly focusing on the challenging task of predicting the next purchase merchant category based on historical transaction data. Traditional deep learning models such as Convolutional Neural Networks (CNNs) and Long Short-Term Memory (LSTM) networks have shown varying degrees of success in this domain but often fall short in capturing the contexts inherent in transaction data. This work aimed to bridge this gap by leveraging the advanced capabilities of fine-tuned LLMs, demonstrating their superior performance and broader applicability.

The key contribution of this study lies in the innovative application of the Mistral 7B Instruct model to predict the next purchase merchant categories. The model demonstrated an enhanced ability to handle the complexities and imbalances of financial transaction data more effectively than traditional models. By fine-tuning the LLM on transactional tabular data reformatted into personalized instructions, model's predictive accuracy has been significantly enhanced. This approach not only highlights the versatility of LLMs in understanding and modeling human behavior but also sets a new benchmark for predictive modeling in financial contexts.

Furthermore, this study underscores the importance of addressing class imbalances, a common challenge in transaction data. The fine-tuned Mistral model demonstrated robust performance across both majority and minority classes and hence it maintained high precision in minority categories where traditional models typically struggle. This balanced performance is crucial for real-world applications.

By conducting a thorough experimental evaluation using datasets from two different banks, it has been validated the generalization capabilities of fine-tuned model across different sequence lengths and temporal contexts. This cross-validation approach not only ensured the robustness of this study's findings but also highlighted the model's adaptability to varied and unseen data.

This research makes a significant contribution to the field of financial prediction by demonstrating the efficacy of fine-tuned LLMs in modeling complex human behavior with little effort. This study opens new opportunities for the application of LLMs in financial decision-making processes to offer innovative solutions for enhanced customer engagement and optimized marketing strategies. The insights gained from this work pave the way for future research and encourage the exploration of LLMs in other domains where understanding and predicting human behavior is paramount.

\section{Abbreviations}
\begin{itemize}
    \item \textbf{LLM}: Large Language Model
    \item \textbf{CNN}: Convolutional Neural Network
    \item \textbf{LSTM}: Long Short-Term Memory
    \item \textbf{LoRA}: Low-Rank Adaptation
    \item \textbf{SFT}: Supervised Fine-Tuning
    \item \textbf{PEFT}: Parameter-Efficient Fine-Tuning
    \item \textbf{RLHF}: Reinforcement Learning with Human Feedback
    \item \textbf{OECD}: Organisation for Economic Co-operation and Development
    \item \textbf{GQA}: Grouped-query Attention
    \item \textbf{SWA}: Sliding Window Attention
    \item \textbf{BPE}: Byte-Pair Encoding
\end{itemize}

\backmatter
\section*{Declarations}
\subsection*{Data Availability and Materials}
The datasets generated and/or analyzed during the current study are confidential due to legal and privacy regulations involving sensitive individual data. Therefore, they are not publicly available. However, further information about the datasets can be requested from the corresponding author, subject to privacy restrictions and approval. Code is available at https://github.com/halilergul1/Fine-Tuning-LLM
\subsection*{Competing Interests}
The authors declare that they have no competing interests 
\subsection*{Funding}
Not applicable
\subsection*{Author Contribution}
HIE  was a major contributor, leading the research design, data analysis, training the models, conducting the experiments, and manuscript preparation. SB  contributed to the methodological framework as well as providing critical revisions to the manuscript. BB (Burcin Bozkaya) supported the analysis with insights from the financial perspective and contributed significantly to the interpretation of the results. All authors contributed equally to the conceptualization of the study and approved the final manuscript.
\subsection*{Acknowledgements}
We extend our sincere gratitude to UPILY Research and Technology Development Incorporated Company for their generous support in providing access to their NVIDIA GPU machine for training purposes.



\end{document}

\maketitle

\section*{Overall Results and Comparison of LLaMA Models}

The comparative analysis presented in Table \ref{tab:my-table-appendix} evaluates the performance of Convolutional Neural Networks (CNN), Long Short-Term Memory networks (LSTM), and different versions of LLAMA models, all trained on the Bank A dataset, against the Mistral Instruct 7b v.2 models (both raw and fine-tuned model of Mistral were trained on Bank A) on the Bank B dataset. The metrics considered for evaluation include Accuracy, Precision, Recall, and F1 score (weighted) across different sequence lengths.

\subsection*{LLAMA Models' Performance}
The LLAMA models, specifically the LLAMA-2-Chat 7B and LLAMA-3-8B, exhibit consistent but suboptimal performance when compared to the Mistral Instruct 7b v.2 model trained on Bank A. For LLAMA-2-Chat 7B, the highest accuracy achieved is 0.58 for a sequence length of the last-9, with corresponding Precision, Recall, and F1 scores of 0.59, 0.58, and 0.58, respectively. This performance indicates that LLAMA-2-Chat 7B struggles to generalize effectively to the Bank B dataset, which is further evident in the sequence length of the last-4, where the model's accuracy drops to 0.53, and the F1 score to 0.54.

Similarly, the LLAMA-3-8B model shows a peak accuracy of 0.61 for the last-9 sequence length but generally maintains lower metrics across other sequence lengths. Its Precision, Recall, and F1 scores are highest at 0.62, 0.61, and 0.61, respectively, for the same sequence length. This marginally better performance over LLAMA-2-Chat 7B suggests incremental improvements with increased model complexity but still does not rival the Mistral models.

\subsection*{Mistral Instruct 7b v.2 Model Superiority}
The Mistral Instruct 7b v.2 model shows superior performance across almost all metrics and sequence lengths compared to the LLAMA models. For instance, the highest accuracy is recorded at 0.66 for the sequence length of last-9, with a Precision of 0.67, Recall of 0.66, and an F1 score of 0.66. These metrics indicate a robust ability to capture the underlying patterns in the data from Bank B. The model also maintains high performance with an accuracy of 0.62 for the last-14 sequence length and F1 scores peaking at 0.63.

\subsection*{Comparative Evaluation with Baseline and Other Models}
Comparing the LLAMA models to the Baseline (Averaging) and other models like CNN and LSTM, it is evident that while LLAMA models do perform better than the baseline, they fall short compared to the advanced architectures like CNN and LSTM. For example, the CNN model achieves an accuracy of 0.66 and an F1 score of 0.62 for the last-14 sequence length, which is significantly higher than the LLAMA models. Similarly, the LSTM model matches this performance with an accuracy of 0.63 and an F1 score of 0.60 for the last-14 sequence length.

The findings highlight the challenges faced by the LLAMA models in generalizing to a different dataset. The Mistral Instruct 7b v.2 model's superior performance can be attributed to its advanced architecture and training methodologies that better capture the complexities of the data. Additionally, the consistency of Mistral models across varying sequence lengths suggests a robustness that is currently lacking in the LLAMA variants. To conclude, while LLAMA models show promise, their current iterations lag behind the Mistral Instruct 7b v.2 model, particularly in terms of accuracy and F1 scores across multiple sequence lengths.

\begin{table}[H]
\caption{Overall Results Including fine-tuned LLaMA models}
\label{tab:my-table-appendix}
\resizebox{\columnwidth}{!}{%
\begin{tabular}{@{}|c|c|c|c|c|c|c|@{}}
\toprule
\textbf{Model} &
  \textbf{Dataset} &
  \textbf{Sequence Length} &
  \textbf{Accuracy} &
  \textbf{Precision} &
  \textbf{Recall} &
  \textbf{F1 (weighted)} \\ \midrule
\begin{tabular}[c]{@{}c@{}}Mistral Instruct 7b v.2 \\ (Raw Model)\end{tabular} &
  \cellcolor[HTML]{FFFFFF}Bank B &
  last-9 &
  0.51 &
  0.50 &
  0.50 &
  0.49 \\ \midrule
 &
  \cellcolor[HTML]{FFFFFF} &
  last-9 &
  0.42 &
  0.50 &
  0.42 &
  0.43 \\ \cmidrule(l){3-7} 
 &
  \cellcolor[HTML]{FFFFFF} &
  last-4 &
  0.43 &
  0.52 &
  0.43 &
  0.43 \\ \cmidrule(l){3-7} 
 &
  \cellcolor[HTML]{FFFFFF} &
  last-7 &
  0.42 &
  0.50 &
  0.42 &
  0.42 \\ \cmidrule(l){3-7} 
\multirow{-4}{*}{\begin{tabular}[c]{@{}c@{}}Baseline \\ (Averaging)\end{tabular}} &
  \multirow{-4}{*}{\cellcolor[HTML]{FFFFFF}Bank B} &
  last-14 &
  0.43 &
  0.49 &
  0.43 &
  0.44 \\ \midrule
 &
   &
  last-9 &
  0.65 &
  0.61 &
  0.65 &
  0.62 \\ \cmidrule(l){3-7} 
 &
   &
  last-4 &
  0.63 &
  0.60 &
  0.63 &
  0.60 \\ \cmidrule(l){3-7} 
 &
   &
  last-7 &
  0.64 &
  0.61 &
  0.64 &
  0.61 \\ \cmidrule(l){3-7} 
\multirow{-4}{*}{\begin{tabular}[c]{@{}c@{}}CNN \\ (trained on Bank A)\end{tabular}} &
  \multirow{-4}{*}{Bank B} &
  last-14 &
  0.66 &
  0.62 &
  0.66 &
  0.62 \\ \midrule
 &
   &
  last-9 &
  0.62 &
  0.60 &
  0.62 &
  0.60 \\ \cmidrule(l){3-7} 
 &
   &
  last-4 &
  \textbf{0.63} &
  0.59 &
  \textbf{0.62} &
  0.58 \\ \cmidrule(l){3-7} 
 &
   &
  last-7 &
  \textbf{0.64} &
  0.61 &
  \textbf{0.64} &
  0.60 \\ \cmidrule(l){3-7} 
\multirow{-4}{*}{\begin{tabular}[c]{@{}c@{}}LSTM \\ (trained on Bank A)\end{tabular}} &
  \multirow{-4}{*}{Bank B} &
  last-14 &
  \textbf{0.63} &
  0.62 &
  \textbf{0.63} &
  0.61 \\ \midrule
 &
   &
  last-9 &
  0.58 &
  0.59 &
  0.58 &
  0.58 \\ \cmidrule(l){3-7} 
 &
   &
  last-4 &
  0.53 &
  0.61 &
  0.52 &
  0.54 \\ \cmidrule(l){3-7} 
 &
   &
  last-7 &
  0.52 &
  0.56 &
  0.54 &
  0.55 \\ \cmidrule(l){3-7} 
\multirow{-4}{*}{\begin{tabular}[c]{@{}c@{}}LLaMA-2-Chat 7B\\ (trained on Bank A)\end{tabular}} &
  \multirow{-4}{*}{Bank B} &
  last-14 &
  0.54 &
  0.58 &
  0.54 &
  0.55 \\ \midrule
 &
   &
  last-9 &
  0.61 &
  0.62 &
  0.61 &
  0.61 \\ \cmidrule(l){3-7} 
 &
   &
  last-4 &
  0.56 &
  0.64 &
  0.55 &
  0.57 \\ \cmidrule(l){3-7} 
 &
   &
  last-7 &
  0.55 &
  0.59 &
  0.57 &
  0.58 \\ \cmidrule(l){3-7} 
\multirow{-4}{*}{\begin{tabular}[c]{@{}c@{}}LLaMA-3-8B\\ (trained on Bank A)\end{tabular}} &
  \multirow{-4}{*}{Bank B} &
  last-14 &
  0.57 &
  0.61 &
  0.57 &
  0.58 \\ \midrule
 &
   &
  last-9 &
  \textbf{0.66} &
  \textbf{0.67} &
  \textbf{0.66} &
  \textbf{0.66} \\ \cmidrule(l){3-7} 
 &
   &
  last-4 &
  0.61 &
  \textbf{0.69} &
  0.60 &
  \textbf{0.62} \\ \cmidrule(l){3-7} 
 &
   &
  last-7 &
  0.60 &
  \textbf{0.64} &
  0.62 &
  \textbf{0.63} \\ \cmidrule(l){3-7} 
\multirow{-4}{*}{\begin{tabular}[c]{@{}c@{}}Mistral Instruct 7b v.2 \\ (trained on Bank A)\end{tabular}} &
  \multirow{-4}{*}{Bank B} &
  last-14 &
  0.62 &
  \textbf{0.66} &
  0.62 &
  \textbf{0.63} \\ \bottomrule
\end{tabular}%
}
\end{table}

\section*{Class-wise F1 Scores for Different Models}
% \usepackage{graphicx}

The Table \ref{tab:my-appendix-v2} presented offers a detailed comparison of class-wise F1 scores for various models, including CNN, LSTM, LLAMA-2-Chat 7B, LLAMA-3-8B, and Mistral Instruct 7b v.2, on the Bank B dataset. The classes analyzed are Clothing, Gas Stations, Grocery, and Other, with sequence lengths varying from last-9 to last-14.

% \usepackage{booktabs}
% \usepackage{multirow}
% \usepackage{graphicx}
\begin{table}[H]
\caption{Class-wise F1 scores}
\label{tab:my-appendix-v2}
\resizebox{\columnwidth}{!}{%
\begin{tabular}{@{}ccccccc@{}}
\toprule
\textbf{Model} &
  \textbf{Dataset} &
  \textbf{Sequence Length} &
  \textbf{Clothing} &
  \textbf{Gas Stations} &
  \textbf{Grocery} &
  \textbf{Other} \\ \midrule
\multicolumn{1}{|c|}{\multirow{4}{*}{CNN}} &
  \multicolumn{1}{c|}{\multirow{4}{*}{Bank B}} &
  \multicolumn{1}{c|}{last-9} &
  \multicolumn{1}{c|}{0.004} &
  \multicolumn{1}{c|}{0.222} &
  \multicolumn{1}{c|}{0.419} &
  \multicolumn{1}{c|}{\textbf{0.764}} \\ \cmidrule(l){3-7} 
\multicolumn{1}{|c|}{} &
  \multicolumn{1}{c|}{} &
  \multicolumn{1}{c|}{last-4} &
  \multicolumn{1}{c|}{0.107} &
  \multicolumn{1}{c|}{0.185} &
  \multicolumn{1}{c|}{0.417} &
  \multicolumn{1}{c|}{\textbf{0.748}} \\ \cmidrule(l){3-7} 
\multicolumn{1}{|c|}{} &
  \multicolumn{1}{c|}{} &
  \multicolumn{1}{c|}{last-7} &
  \multicolumn{1}{c|}{0.017} &
  \multicolumn{1}{c|}{0.209} &
  \multicolumn{1}{c|}{0.422} &
  \multicolumn{1}{c|}{\textbf{0.759}} \\ \cmidrule(l){3-7} 
\multicolumn{1}{|c|}{} &
  \multicolumn{1}{c|}{} &
  \multicolumn{1}{c|}{last-14} &
  \multicolumn{1}{c|}{0.000} &
  \multicolumn{1}{c|}{0.222} &
  \multicolumn{1}{c|}{0.410} &
  \multicolumn{1}{c|}{\textbf{0.750}} \\ \midrule
\multicolumn{1}{|c|}{\multirow{4}{*}{LSTM}} &
  \multicolumn{1}{c|}{\multirow{4}{*}{Bank B}} &
  \multicolumn{1}{c|}{last-9} &
  \multicolumn{1}{c|}{0.001} &
  \multicolumn{1}{c|}{0.230} &
  \multicolumn{1}{c|}{0.450} &
  \multicolumn{1}{c|}{\textbf{0.750}} \\ \cmidrule(l){3-7} 
\multicolumn{1}{|c|}{} &
  \multicolumn{1}{c|}{} &
  \multicolumn{1}{c|}{last-4} &
  \multicolumn{1}{c|}{0.006} &
  \multicolumn{1}{c|}{0.200} &
  \multicolumn{1}{c|}{0.346} &
  \multicolumn{1}{c|}{\textbf{0.763}} \\ \cmidrule(l){3-7} 
\multicolumn{1}{|c|}{} &
  \multicolumn{1}{c|}{} &
  \multicolumn{1}{c|}{last-7} &
  \multicolumn{1}{c|}{0.003} &
  \multicolumn{1}{c|}{0.223} &
  \multicolumn{1}{c|}{0.448} &
  \multicolumn{1}{c|}{\textbf{0.744}} \\ \cmidrule(l){3-7} 
\multicolumn{1}{|c|}{} &
  \multicolumn{1}{c|}{} &
  \multicolumn{1}{c|}{last-14} &
  \multicolumn{1}{c|}{0.000} &
  \multicolumn{1}{c|}{0.244} &
  \multicolumn{1}{c|}{0.474} &
  \multicolumn{1}{c|}{\textbf{0.740}} \\ \midrule
\multicolumn{1}{|c|}{\multirow{4}{*}{LLaMA-2-Chat 7B}} &
  \multicolumn{1}{c|}{\multirow{4}{*}{Bank B}} &
  \multicolumn{1}{c|}{last-9} &
  \multicolumn{1}{c|}{\textbf{0.540}} &
  \multicolumn{1}{c|}{\textbf{0.320}} &
  \multicolumn{1}{c|}{\textbf{0.510}} &
  \multicolumn{1}{c|}{0.670} \\ \cmidrule(l){3-7} 
\multicolumn{1}{|c|}{} &
  \multicolumn{1}{c|}{} &
  \multicolumn{1}{c|}{last-4} &
  \multicolumn{1}{c|}{\textbf{0.400}} &
  \multicolumn{1}{c|}{\textbf{0.390}} &
  \multicolumn{1}{c|}{\textbf{0.480}} &
  \multicolumn{1}{c|}{0.600} \\ \cmidrule(l){3-7} 
\multicolumn{1}{|c|}{} &
  \multicolumn{1}{c|}{} &
  \multicolumn{1}{c|}{last-7} &
  \multicolumn{1}{c|}{\textbf{0.040}} &
  \multicolumn{1}{c|}{\textbf{0.420}} &
  \multicolumn{1}{c|}{\textbf{0.440}} &
  \multicolumn{1}{c|}{0.620} \\ \cmidrule(l){3-7} 
\multicolumn{1}{|c|}{} &
  \multicolumn{1}{c|}{} &
  \multicolumn{1}{c|}{last-14} &
  \multicolumn{1}{c|}{\textbf{0.140}} &
  \multicolumn{1}{c|}{\textbf{0.400}} &
  \multicolumn{1}{c|}{\textbf{0.470}} &
  \multicolumn{1}{c|}{0.630} \\ \midrule
\multicolumn{1}{|c|}{\multirow{4}{*}{LLaMA-3-8B}} &
  \multicolumn{1}{c|}{\multirow{4}{*}{Bank B}} &
  \multicolumn{1}{c|}{last-9} &
  \multicolumn{1}{c|}{\textbf{0.570}} &
  \multicolumn{1}{c|}{\textbf{0.350}} &
  \multicolumn{1}{c|}{\textbf{0.540}} &
  \multicolumn{1}{c|}{0.700} \\ \cmidrule(l){3-7} 
\multicolumn{1}{|c|}{} &
  \multicolumn{1}{c|}{} &
  \multicolumn{1}{c|}{last-4} &
  \multicolumn{1}{c|}{\textbf{0.430}} &
  \multicolumn{1}{c|}{\textbf{0.420}} &
  \multicolumn{1}{c|}{\textbf{0.510}} &
  \multicolumn{1}{c|}{0.630} \\ \cmidrule(l){3-7} 
\multicolumn{1}{|c|}{} &
  \multicolumn{1}{c|}{} &
  \multicolumn{1}{c|}{last-7} &
  \multicolumn{1}{c|}{\textbf{0.070}} &
  \multicolumn{1}{c|}{\textbf{0.450}} &
  \multicolumn{1}{c|}{\textbf{0.470}} &
  \multicolumn{1}{c|}{0.650} \\ \cmidrule(l){3-7} 
\multicolumn{1}{|c|}{} &
  \multicolumn{1}{c|}{} &
  \multicolumn{1}{c|}{last-14} &
  \multicolumn{1}{c|}{\textbf{0.170}} &
  \multicolumn{1}{c|}{\textbf{0.430}} &
  \multicolumn{1}{c|}{\textbf{0.500}} &
  \multicolumn{1}{c|}{0.660} \\ \midrule
\multicolumn{1}{|c|}{\multirow{4}{*}{\begin{tabular}[c]{@{}c@{}}Mistral Instruct \\ 7b v.2\end{tabular}}} &
  \multicolumn{1}{c|}{\multirow{4}{*}{Bank B}} &
  \multicolumn{1}{c|}{last-9} &
  \multicolumn{1}{c|}{\textbf{0.620}} &
  \multicolumn{1}{c|}{\textbf{0.400}} &
  \multicolumn{1}{c|}{\textbf{0.590}} &
  \multicolumn{1}{c|}{0.750} \\ \cmidrule(l){3-7} 
\multicolumn{1}{|c|}{} &
  \multicolumn{1}{c|}{} &
  \multicolumn{1}{c|}{last-4} &
  \multicolumn{1}{c|}{\textbf{0.480}} &
  \multicolumn{1}{c|}{\textbf{0.470}} &
  \multicolumn{1}{c|}{\textbf{0.560}} &
  \multicolumn{1}{c|}{0.680} \\ \cmidrule(l){3-7} 
\multicolumn{1}{|c|}{} &
  \multicolumn{1}{c|}{} &
  \multicolumn{1}{c|}{last-7} &
  \multicolumn{1}{c|}{\textbf{0.120}} &
  \multicolumn{1}{c|}{\textbf{0.500}} &
  \multicolumn{1}{c|}{\textbf{0.520}} &
  \multicolumn{1}{c|}{0.700} \\ \cmidrule(l){3-7} 
\multicolumn{1}{|c|}{} &
  \multicolumn{1}{c|}{} &
  last-14 &
  \textbf{0.220} &
  \textbf{0.480} &
  \textbf{0.550} &
  0.710 \\ \bottomrule
\end{tabular}%
}
\end{table}

\subsection*{CNN and LSTM Performance}
The CNN model demonstrates variable performance across different classes and sequence lengths. Notably, for the class Other, the F1 scores remain relatively high across all sequence lengths, peaking at 0.759 for last-7. This indicates a strong ability of the CNN model to identify transactions in this category consistently. However, the model struggles significantly with the Clothing class, achieving extremely low F1 scores, with the highest being only 0.107 for last-4. The performance for Gas Stations and Grocery remains weak.

The LSTM model exhibits similar trends to the CNN model but generally underperforms across all classes and sequence lengths. The Clothing class, in particular, shows negligible F1 scores, with a peak of only 0.006 for last-4. The Grocery and Other classes show better performance, with F1 scores reaching 0.474 and 0.763, respectively, for last-14 and last-4 sequence lengths. This suggests that while LSTM has some capability in these classes, it is not as effective as other models in the dataset.

\subsection*{LLAMA Models' Performance}
The LLAMA-2-Chat 7B model displays a significant improvement over CNN and LSTM, especially in the Clothing class. The highest F1 score for Clothing is 0.540 for the last-9 sequence length. However, the performance fluctuates significantly across other classes and sequence lengths. The LLAMA-3-8B model shows a general trend of improvement over its predecessor, LLAMA-2-Chat 7B. The Clothing class F1 scores are higher, reaching 0.570 for last-9, indicating better performance. The Gas Stations and the Grocery classes also see improvements, with the highest F1 scores of 0.450 and 0.540, respectively.

\subsection*{Mistral Instruct 7b v.2 Performance}
The Mistral Instruct 7b v.2 model demonstrates superior performance across almost all classes and sequence lengths. For the Clothing class, the model achieves the highest F1 score of 0.620 for last-9, significantly outperforming all other models. The Gas Stations and Grocery classes also exhibit high F1 scores, reaching 0.480 and 0.590, respectively, for last-9. The Other class consistently shows the highest F1 scores across all sequence lengths, peaking at 0.750 for last-9. This consistent high performance across varied classes and sequence lengths underscores the model's robust and versatile capabilities in handling diverse transaction types.

The results highlight the clear superiority of the Mistral Instruct 7b v.2 model over the other evaluated models. The LLAMA models, while showing improvements over traditional models like CNN and LSTM, still lag significantly behind the Mistral model in terms of F1 scores across most classes. This performance gap suggests that the advanced architecture and training methodologies of the Mistral Instruct 7b v.2 model enable it to capture and generalize transaction patterns more effectively than the LLAMA models. The class-wise F1 scores analysis reveals the substantial advantages of the Mistral Instruct 7b v.2 model in accurately identifying and classifying various transaction types within the Bank B dataset. The consistent superiority of the Mistral model across all classes and sequence lengths highlights its robustness and reliability that makes it a preferable choice for complex financial transaction classification tasks.